\documentclass[10pt,english,aps,manuscript]{revtex4}
\usepackage[T1]{fontenc}
\usepackage[latin9]{inputenc}
\usepackage{amsmath}
\usepackage{amssymb}
\usepackage{graphicx}

\makeatletter
\@ifundefined{textcolor}{}
{%
 \definecolor{BLACK}{gray}{0}
 \definecolor{WHITE}{gray}{1}
 \definecolor{RED}{rgb}{1,0,0}
 \definecolor{GREEN}{rgb}{0,1,0}
 \definecolor{BLUE}{rgb}{0,0,1}
 \definecolor{CYAN}{cmyk}{1,0,0,0}
 \definecolor{MAGENTA}{cmyk}{0,1,0,0}
 \definecolor{YELLOW}{cmyk}{0,0,1,0}
 }

\@ifundefined{textcolor}{}{%
 \definecolor{BLACK}{gray}{0}
 \definecolor{WHITE}{gray}{1}
 \definecolor{RED}{rgb}{1,0,0}
 \definecolor{GREEN}{rgb}{0,1,0}
 \definecolor{BLUE}{rgb}{0,0,1}
 \definecolor{CYAN}{cmyk}{1,0,0,0}
 \definecolor{MAGENTA}{cmyk}{0,1,0,0}
 \definecolor{YELLOW}{cmyk}{0,0,1,0}
 }

\usepackage{paralist}
\def\btt#1{\texttt{\@backslashchar#1}}
\DeclareRobustCommand{\bblash}{\btt{\@backslashchar}}

\usepackage{babel}

\makeatother

\usepackage{babel}
\begin{document}

\title{Phase Diagram of the mixed spin-2 and spin-5/2 Ising system with
two different single-ion anisotropies }

\author{J. S. da Cruz Filho}

\email{zefilho@fisica.ufmt.br}

\selectlanguage{english}%

\author{M. Godoy}

\email{mgodoy@fisica.ufmt.br}

\selectlanguage{english}%

\author{A. S. de Arruda}

\email{aarruda@fisica.ufmt.br}

\selectlanguage{english}%

\affiliation{Instituto de Física, Universidade Federal de Mato Grosso, 78060-900,
Cuiabá, Mato Grosso, Brazil.}
\begin{abstract}
In this paper we present a study of the effects of two different single-ion
anisotropies in the phase diagram and in the compensation temperature
of the mixed spin-2 and spin-5/2 Ising ferrimagnetic system. We employed
the mean-field theory based on the Bogoliubov inequality for the Gibbs
free energy. Also we found the Landau expansion of the free energy
in the order parameter to describe the phase diagrams. In the plane
critical temperature versus single-ion anisotropie the phase diagram
displays tricritical behavior. The critical and compesation temperatures
increase when the single-ion anisotropies increase. 
\end{abstract}
\maketitle

\section{Introduction}

In the last five decades the Ising model has been one of the most
largely used to describe critical behavior of several systems in the
nature. In particular, in the physics of the condensed matter it is
important to describe critical behavior and other thermodynamics properties
of a variety of physic systems (disordered system, spins glass, random
field Ising model, etc.). Recently, several extensions have been made
in the spin$-1/2$ Ising model to describe a wide variety of physic
systems. For example, the models consisting of mixed spins of different
magnitudes are interesting extensions, which are the so-called mixed-spin
Ising models. 

The researches in ferrimagnetic materials are of great interest due
to their possible technological applications, as well as of the academic
point of view. These materials are modeled by mixed-spin Ising model,
which can be constituted by several combinations of spins (spin-1/2,
spin-1), (spin-1/2, spin-3/2), (spin-1, spin-3/2), (spin-2, spin-3/2),
(spin-2, spin-5/2). The interest in studying magnetic properties of
some types of ferrimagnetism, namely the molecular-based magnetic
materials \cite{Kahn,Kaneyoshi1,Kaneyoshi2}, is due to its less translational
symmetry than to their single-spin counterparts since they consist
of two interpenetrating sublattices. The bimetallic chain complex
$MnNi(EDTA)-6H_{2}0$ is an example of an experimental mixed-spin
system \cite{Drillon}. 

Additionally, there are many studies of mixed-spin Ising systems for
explanation of the physical properties of disordered systems. This
theme has been a great challenge in statistical mechanics. In this
sense, in the last years there has been great interest in the study
of magnetic properties of systems formed by two sublattices with different
spins and with different crystal field interactions\cite{Bobak}.

One of the earliest and simplest of these models to be studied was
the mixed-spin Ising system consisting of spin-1/2 and spin-$S$ ($S>1/2$)
in an uniaxial crystal field \cite{Resende,Keskin}. From the theoretical
point of view, such systems have been widely studied by a variety
of approaches, for instance, effective-field theory \cite{Kaneyoshi0,Kaneyoshi3,Kaneyoshi4,Benyoussef1,Benyoussef2,Bobak2,Arruda1,Arruda2,mohamad},
mean-field approximation \cite{Kaneyoshi5}, renormalization-group
technique \cite{Salinas} and Monte-Carlo simulation \cite{Zhang,Buendia1,Buendia2,godoy,gu,wei,zukovic,zukovic-1,yessoufou}
and exact solutions for the mixed spin-1 and spin-S Ising model in
an uniaxial crystal field \cite{ll1,ll2}.

Recently, the goal is to extend the investigations to a more general
mixed-spin Ising model with one constituent spin-1 and, in the simplest
case, the other constituent spin-3/2. Abubrig \emph{et al.} \cite{Abubrig}
presented a study of mean-field theory to determine the effects in
the phase diagram of different crystal fields of the mixed spin-1
and spin-3/2 Ising system. They also showed some outstanding features
in the temperature dependence of the total and sublattice magnetization.

In this paper, we studied the effects of two different single-ion
anisotropies in the phase diagram and in the compensation temperature
of the mixed spin-2 and spin-5/2 Ising ferrimagnetic system. The outline
of the remainder of the paper is as follows: In Section II, the model
is introduced and we have obtained analytical expressions for free-energy,
equations of state. We also have found the Landau expansion of the
free energy in the order parameter. In Section III, we present the
results and discussions about the phase diagrams and the compensation
temperature. Finally, in Section IV we present our conclusions.

\section{The model and Calculation}

The mixed-spin ferrimagnetic Ising system consists of two interpenetrating
square sublattices (A and B) with spin $S^{A}=0,\pm1,\pm2$ and spin
$S^{B}=\pm1/2,\pm3/2,\pm5/2$. In each site of the lattice there is
a single-ion anisotropie ($D_{A}$ in the sublattice A and $D_{B}$
in the sublattice B) acting in the spin-2 and spin-5/2. The system
is described by the following Hamiltonian model: 
\begin{eqnarray}
{\cal H}=-J\sum_{\left<i,j\right>}S_{i}^{A}S_{j}^{B}-D_{A}\sum_{i\in A}(S_{i}^{A})^{2}-D_{B}\sum_{j\in B}(S_{j}^{B})^{2},\label{eq:1}
\end{eqnarray}
where the first term represents interaction between the nearest neighbors
spins in the sites $i$ and $j$ located in the different sublattices
$A$ and $B$. $J$ is the magnitude of this interaction, and the
sum is over all nearest neighboring pairs of spins. The second and
third terms represent the single-ion anisotropies at all points of
the sublattices $A$ and $B$, respectively. The sums are performed
on $N/2$ spins of each sublattice. 

In order to derive the analytical expressions for free-energy and
equations of state, we employed the variational method based on the
Bogoliubov inequality for the Gibbs free energy 
\begin{equation}
G({\cal H})\leq G_{0}({\cal H}_{0})+\langle{\cal H}-{\cal H}_{0}(\eta)\rangle_{0}=\Phi(\eta),\label{eq:2}
\end{equation}
where $G({\cal H})$ is the free energy of ${\cal H}$, and $G_{0}({\cal H}_{0})$
is the free energy of a trial Hamiltonian ${\cal H}_{0}(\eta)$ depending
on variational parameters. $\langle\cdots\rangle_{0}$ denotes a thermal
average over the ensemble defined by ${\cal H}_{0}(\eta)$. To facilitate
the calculations, we choose the simplest trial Hamiltonian, which
is given by 
\begin{eqnarray}
{\cal H}_{0} & = & -\sum_{i\in A}\left[D_{A}(S_{i}^{A})^{2}+\eta_{A}S_{i}^{A}\right]-\sum_{j\in B}\left[D_{B}(S_{j}^{B})^{2}+\eta_{B}S_{j}^{B}\right],\label{eq:3}
\end{eqnarray}
where $\eta_{A}$ and $\eta_{B}$ are variational parameters related
to the two different spins. Through this approach, we found the free
energy and the equations of state (sublattice magnetization per site
$m_{A}$ and $m_{B}$): 
\begin{eqnarray}
g=\frac{\phi}{N} & = & -\frac{1}{2\beta}\ln\left[2\exp{(4\beta D_{A})}\cosh{(2\beta\eta_{A})}+2\exp{(\beta D_{A})}\cosh{(\beta\eta_{A})}+1\right]\nonumber \\
 &  & -\frac{1}{2\beta}\ln\bigg[2\exp{\bigg(\frac{25}{4}\beta D_{B}\bigg)}\cosh{\bigg(\frac{5}{2}\beta\eta_{B}\bigg)}+2\exp{\bigg(\frac{9}{4}\beta D_{B}\bigg)}\cosh{\bigg(\frac{3}{2}\beta\eta_{B}\bigg)}\nonumber \\
 &  & +2\exp{\bigg(\frac{1}{4}\beta D_{B}\bigg)}\cosh{\bigg(\frac{1}{2}\beta\eta_{B}\bigg)}\bigg]-\frac{1}{2}Jzm_{A}m_{B}+\frac{1}{2}\eta_{A}m_{A}+\frac{1}{2}\eta_{B}m_{B},\label{eq:4}
\end{eqnarray}
where $g=\frac{\phi}{N}$, $\beta=1/k_{B}T$, $N$ is the total number
of sites of the lattice and $z$ is the coordination number. The sublattice
magnetization per site $m_{A}$ and $m_{B}$ are defined by $m_{A}=\langle S_{i}^{A}\rangle_{0}$
and $m_{B}=\langle S_{j}^{B}\rangle_{0}$, thus 
\begin{equation}
m_{A}=\frac{2\sinh{(2\beta\eta_{A})}+\exp{(-3\beta D_{A})}\sinh{(\beta\eta_{A})}}{\cosh{(2\beta\eta_{A})}+\exp{(-3\beta D_{A})}\cosh{(\beta\eta_{A})}+\frac{1}{2}\exp{(-4\beta D_{A})}},\label{eq:5}
\end{equation}
and 
\begin{equation}
m_{B}=\frac{1}{2}\left[\frac{\exp{(-6\beta D_{B})}\sinh{(\frac{1}{2}\beta\eta_{B})}+3\exp{(-4\beta D_{B})}\sinh{(\frac{3}{2}\beta\eta_{B})}+5\sinh{(\frac{5}{2}\beta\eta_{B})}}{\exp{(-6\beta D_{B})}\cosh{(\frac{1}{2}\beta\eta_{B})}+\exp{-4\beta D_{B}}\cosh{(\frac{3}{2}\beta\eta_{B})}+\cosh{(\frac{5}{2}\beta\eta_{B})}}\right].\label{eq:6}
\end{equation}
Minimizing the free energy in terms of the variational parameters
$\eta_{A}$ and $\eta_{B}$, we have obtained 
\begin{eqnarray}
\eta_{A} & = & Jzm_{B}\,\,\,,\,\,\,\eta_{B}=Jzm_{A}.\label{eq:7}
\end{eqnarray}
Therefore, we can get the generic mean-field equations (\ref{eq:4}-\ref{eq:7}),
which provide the magnetic properties of ferrimanetic system. Since
these equations (\ref{eq:5}-\ref{eq:7}) have in general several
solutions for the pair ($m_{A}$ and $m_{B}$), the stable phase will
be one which minimizes the free energy. The detailed phase diagram
is determined by numerical analysis, but some features of the phases
diagram can be obtained analytically. Therefore, close to the second-order
phase transition from an ordered state ($m_{A}\neq$ and $m_{B}\neq0$)
to a disordered state ($m_{A}=0$ and $m_{B}=0$). The magnetization
$m_{A}$ and $m_{B}$ are very small, so we can expand the equations
(\ref{eq:4}-\ref{eq:6}) to obtain a Landau-like expansion: 
\begin{equation}
g=A_{0}+A_{2}m_{A}^{2}+A_{4}m_{A}^{4}+A_{6}m_{A}^{6}+\cdots,\label{eq:8}
\end{equation}
where the expansion coefficients are given by 
\begin{equation}
A_{0}=-\frac{1}{2\beta}\ln[(1+y_{a}+x_{a})(z_{b}+y_{b}+x_{b})],\label{eq:9}
\end{equation}
 
\begin{equation}
A_{2}=\frac{1}{2\beta}\bigg[\frac{t^{2}}{4}a_{1}-\frac{t^{2}}{8}a_{2}-\frac{t^{4}}{32}a_{1}^{2}b_{1}\bigg],\label{eq:10}
\end{equation}
 
\begin{eqnarray}
A_{4} & = & \frac{1}{2\beta}\biggl[\frac{t^{4}}{768}a_{1}^{2}c_{1}+\frac{t^{3}}{192}c_{2}a_{1}a_{2}+\frac{t^{2}}{96}c_{3}\biggr],\label{eq:11}
\end{eqnarray}
 
\begin{eqnarray}
A_{6} & = & \frac{1}{2\beta}\bigg[\frac{t^{4}}{11520}\bigg(c_{4}+\frac{6}{t}c_{5}\bigg)+t^{5}\bigg(\frac{2a_{2}}{18423}c_{2}(3a_{1}^{2}-a_{3})-\frac{c_{5}}{7680}a_{2}a_{1}\bigg)+\frac{t^{7}c_{2}}{18432}\bigg(a_{4}-3a_{1}^{3}a_{2}^{2}\bigg)+\frac{t^{8}}{245760}c_{6}\bigg],\label{eq:12}
\end{eqnarray}
with $t=\beta Jz$, and 
\begin{eqnarray}
x_{a} & = & 2e^{4\beta D_{A}}\,;\, y_{a}=2e^{\beta D_{A}}\,;\, z_{b2}=e^{-6\beta D_{B}}\,;\, z_{b1}=e^{-4\beta D_{B}},\nonumber \\
x_{b} & = & 2e^{\frac{25}{4}\beta D_{B}};\; y_{b}=2e^{\frac{9}{4}\beta D_{B}};\; z_{b}=2e^{\frac{1}{4}\beta D_{B}}\,;\, t=zJ\beta,\label{eq:13 }
\end{eqnarray}
 
\[
a_{1}=\frac{z_{b_{2}}+9z_{b_{1}}+25}{z_{b_{2}}+z_{b_{1}}+1},\, a_{2}=\frac{4x_{a}+y_{a}}{x_{a}+y_{a}+1},\, b_{1}=\frac{25x_{b}+9y_{b}+z_{b}}{x_{b}+y_{b}+z_{b}},\,
\]
 
\begin{equation}
a_{3}=\frac{z_{b_{2}}+81z_{b_{1}}+625}{z_{b_{2}}+z_{b_{1}}+1},\, a_{4}=\frac{16x_{a}+y_{a}}{x_{a}+y_{a}+1},\, b_{2}=\frac{625x_{b}+81y_{b}+z_{b}}{x_{b}+y_{b}+z_{b}},\,\label{eq:23}
\end{equation}
 
\begin{equation}
a_{5}=\frac{z_{b_{2}}+729z_{b_{1}}+15625}{z_{b_{2}}+z_{b_{1}}+1},\, a_{6}=\frac{64x_{a}+y_{a}}{x_{a}+y_{a}+1},\, b_{3}=\frac{15625x_{b}+729y_{b}+z_{b}}{x_{b}+y_{b}+z_{b}},\,\label{eq:24}
\end{equation}
 
\begin{equation}
c_{1}=\frac{t^{4}}{8}a_{1}^{2}(3a_{2}^{2}-a_{4}),\, c_{2}=\frac{t^{3}}{2}(3a_{1}^{2}-a_{3}),\, c_{3}=\frac{t^{2}}{4}(3b_{1}^{2}+4a_{3}-12a_{1}^{2}-b_{2}),\,\label{eq:25}
\end{equation}
 
\begin{equation}
c_{4}=\frac{t^{2}}{4}(-b_{3}+15b_{1}b_{2}-30-b_{1}),\, c_{5}=\frac{t^{3}}{4}(-15a_{3}a_{1}+a_{5}+30a_{1}^{3}),\,\label{eq:26}
\end{equation}
 
\begin{equation}
c_{6}=\frac{t^{4}}{12}(-a_{6}a_{1}^{6}-30(a_{2}a_{1}^{2})^{3}+15a_{2}a_{4}a_{1}^{6}).\label{eq:27}
\end{equation}

\section{Results and discussions}

The phase diagrams were constructed according to the following routine:
i) numerical solutions of $A_{2}=0$ and $A_{4}>0$ provides second-order
transition lines. ii) $A_{2}=0$, $A_{4}=0$ and $A_{6}>0$ determines
the tricritical points. iii) The first-order transition lines are
determined by comparing the corresponding Gibbs free energies of the
various solutions of equations (\ref{eq:5}) and (\ref{eq:6}) for
the pair ($m_{A}$, $m_{B}$). Even so, we have also analysed that
$A_{6}>0$ in all $T,D_{A},D_{B}$ space. 

The particular case $D_{A}=D_{B}=0$, the critical temperature is
determined by taking $m_{A}\rightarrow0$ and $m_{B}\rightarrow0$,
or $A_{2}=0$ and $D_{A}=D_{B}=0$, thus $k_{B}T_{c}/J=9.6609$.

\subsection{The ground-state}

The ground-state phase diagram (see Fig. 1) is determined from the
Hamiltonian (\ref{eq:1}) by comparing the ground-state energies of
the different phases. At zero temperature, the structure of the ground
state of the system consists of six phases with different values of
$\{m_{A},m_{B},q_{A},q_{B}\}$, namely the ordered ferrimagnetic phases
\[
O_{1}=\bigg\{-2,\frac{5}{2},4,\frac{25}{4}\bigg\},\;\;\;\; O_{2}=\bigg\{-1,\frac{5}{2},1,\frac{25}{4}\bigg\},
\]

\[
O_{3}=\bigg\{-2,\frac{3}{2},4,\frac{9}{4}\bigg\},\;\;\;\; O_{4}=\bigg\{-1,\frac{3}{2},1,\frac{9}{4}\bigg\},
\]

\[
O_{5}=\bigg\{-2,\frac{1}{2},4,\frac{1}{4}\bigg\},\;\;\;\; O_{6}=\bigg\{-1,\frac{1}{2},1,\frac{1}{4}\bigg\},
\]
where $q_{A}=\langle(S_{i}^{A})^{2}\rangle$ and $q_{B}=\langle(S_{i}^{B})^{2}\rangle$.
The energies are given by

\[
E_{1}=-\bigg(\frac{5}{2}Jz+2D_{A}+\frac{25}{8}D_{B}\bigg),\;\;\;\; E_{2}=-\bigg(\frac{5}{4}Jz+\frac{1}{2}D_{A}+\frac{25}{8}D_{B}\bigg),
\]

\[
E_{3}=-\bigg(\frac{3}{2}Jz+2D_{A}+\frac{9}{8}D_{B}\bigg),\;\;\;\; E_{4}=-\bigg(\frac{3}{4}Jz+\frac{1}{2}D_{A}+\frac{9}{8}D_{B}\bigg),
\]

\[
E_{5}=-\bigg(\frac{1}{2}Jz+2D_{A}+\frac{1}{8}D_{B}\bigg),\;\;\;\; E_{6}=-\bigg(\frac{1}{4}Jz+\frac{1}{2}D_{A}+\frac{1}{8}D_{B}\bigg).
\]

\begin{figure}[h]
\centering{} \includegraphics[clip,scale=0.5]{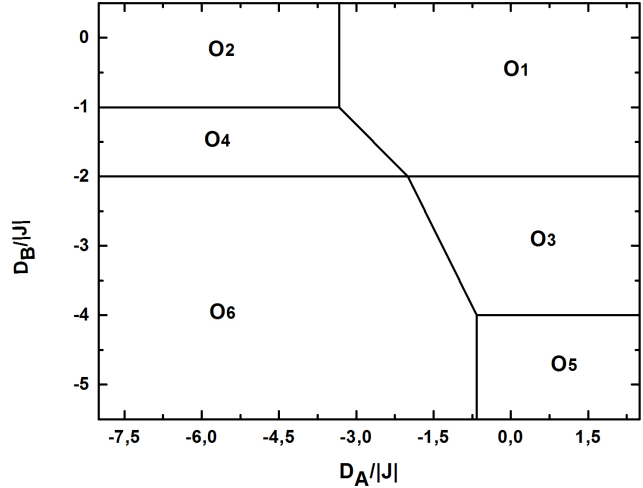} \caption{Ground-state phase diagram of the mixed spin-2 and spin-5/2 Ising
ferrimagnetic system with two different single-ion anisotropies $D_{A}/\left|J\right|$
and $D_{B}/\left|J\right|$. The six ordered phases are represents
by $O_{1}\{-2,5/2,4,25/4\}$, $O_{2}\{-1,5/2,1,25/4\}$, $O_{3}\{-2,3/2,4,9/4\}$,
$O_{4}\{-1,3/2,4,9/4\}$, $O_{5}\{-2,1/2,4,1/4\}$, $O_{6}\{-1,1/2,1,1/4\}$.}
\end{figure}

\subsection{Phase diagram}

The phase diagrams are analysed in the $(D_{A}/\left|J\right|,k_{B}T/\vert J\vert)$
and $(D_{B}/\left|J\right|,k_{B}T/\vert J\vert)$ planes, where the
numerical solutions of $A_{2}=0$ with $A_{4}>0$ provides second-order
transition lines. The tricritical points are determined by numerical
solutions of $A_{2}=0$, $A_{4}=0$ with $A_{6}>0$. The first-order
transition lines were determined by comparing the corresponding Gibbs
free energies of the various solutions of equations (\ref{eq:5})
and (\ref{eq:6}) for the pair ($m_{A}$, $m_{B}$). 

In Fig. 2, we display the phase diagram of $k_{B}T_{c}/|J|$ versus
$D_{A}/\vert J\vert$ for selected values of $D_{B}/\vert J\vert$.
In regions of high temperatures, for all positives and negatives values
$D_{A}/\vert J\vert$, and for any values of $D_{B}/\vert J\vert$,
the phase diagram shows only second-order phase transitions, which
are indicated by solid lines. For values of $D_{B}/\vert J\vert>10.0$,
all second-order lines end in the same tricritical point given by
$(k_{B}T_{t}/|J|=4.5254,D_{A}/\vert J\vert=-9.3200)$. However, for
values of $D^{B}/\vert J\vert<-10.0$, now in low temperatures, also
all second-order lines end in same tricritical point given by $(k_{B}T_{c}/|J|=0.8972,D_{A}/\vert J\vert=-1.8667)$.
One heavy dotted curve connects these two tricritical points $(k_{B}T_{c}/|J|=4.5254,D_{A}/\vert J\vert=-9.3200)$
and $(k_{B}T_{c}/|J|=0.89716,D_{A}/\vert J\vert=-1.8667)$. This curve
separates the region with second-order phase transition of the region
with first-order phase transition. In the region of low temperatures
and for all values of $D_{A}/\vert J\vert$ and $D_{B}/\vert J\vert$,
the phase transitions are of first-order. Thus, in this space $(D_{A}/\left|J\right|,k_{B}T/\vert J\vert)$,
the system presents tricritical behavior. 

Additionally, the diagram shows that when $D_{B}/\left|J\right|\rightarrow+\infty$,
the mixed spin Ising system behaves like a two-level system since
the spin-5/2 behaves like $S^{B}=\pm5/2$. Nevertheless, in case that
$D_{B}/\left|J\right|\rightarrow-\infty$, the $S^{B}=\pm5/2$ states
are suppressed and the system becomes equivalent to a mixed spin-1/2
and spin-2 Ising model. So, this is the reason that the coordinates
of the tricritical point in the limit of large positive $D_{B}/\left|J\right|$
are well higher than those for large negative $D_{B}/\left|J\right|$.
For the special case with equal anisotropic fields $(D_{A}/\left|J\right|=D_{B}/\left|J\right|=0)$,
the critical temperature is $k_{B}T_{c}/\vert J\vert=9.6609$, and
the location of the tricritical point is $k_{B}T_{t}/\vert J\vert=3.3833,D_{A}/\vert J\vert=-6.2150$,
for $z=4$.

\begin{figure}[h]
\centering{} \includegraphics[clip,scale=0.54]{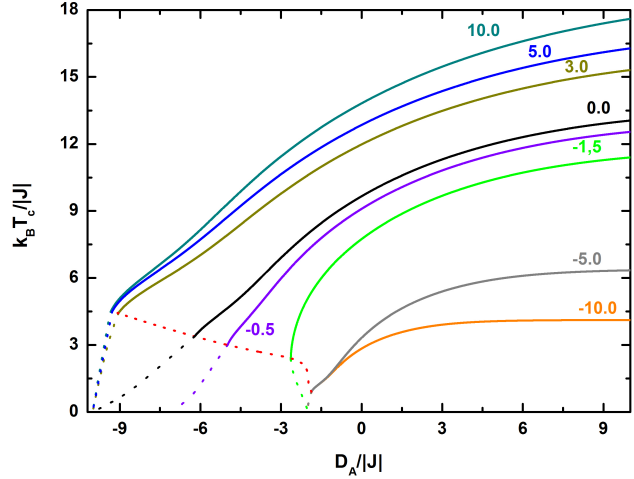} \caption{Phase diagram in the ($D_{A}/\vert J\vert,k_{B}T/\vert J\vert$) plane
for the mixed-spin Ising ferrimagnet with the coordination number
$z=4$, for several values of $D_{B}/\vert J\vert$. The solid and
light dashed lines, respectively, indicate second- and first-order
phase transitions, while the heavy dashed line represents the positions
of tricritical points. }
\end{figure}

In Fig. 3, it is shown the phase diagram of $k_{B}T_{c}/|J|$ versus
$D_{B}/\vert J\vert$ for various values of $D_{A}/\vert J\vert$.
In the case of $D_{A}/\vert J\vert>-1.8842$ the phase transitions
are only of second-order (solid lines) for any values of $D_{B}/\vert J\vert$.
The value of the critical temperature increases when $D_{B}/\vert J\vert$
and $D_{A}/\vert J\vert$ also increase. Still, in range $-1.8842<D_{A}/\vert J\vert<-9.0$,
the phase transitions are of second-order in the high temperatures
region (solid lines) and of first-order (light dashed lines) in the
low temperatures region. One heavy dotted curve of tricritical points
separates the second from the first-order transition lines. 

One interesting feature shown in the diagram of the Fig. 3, refers
to the fact that the phase transitions are only of first-order when
the values of $D_{A}/\vert J\vert<-9.0$. All lines that start below
the heavy dotted line will necessarily be only of first-order phase
transition. Again, in this space $(D_{B}/\vert J\vert,k_{B}T/\vert J\vert)$,
the system presents tricritical behavior. 

These results may be compared to those obtained in the paper\cite{mohamad},
which it has an error in equation (4) and consequently in the equation
(8). These errors led to very different phase diagrams at temperatures
above zero. 

\begin{figure}[h]
\centering{} \includegraphics[clip,scale=0.55]{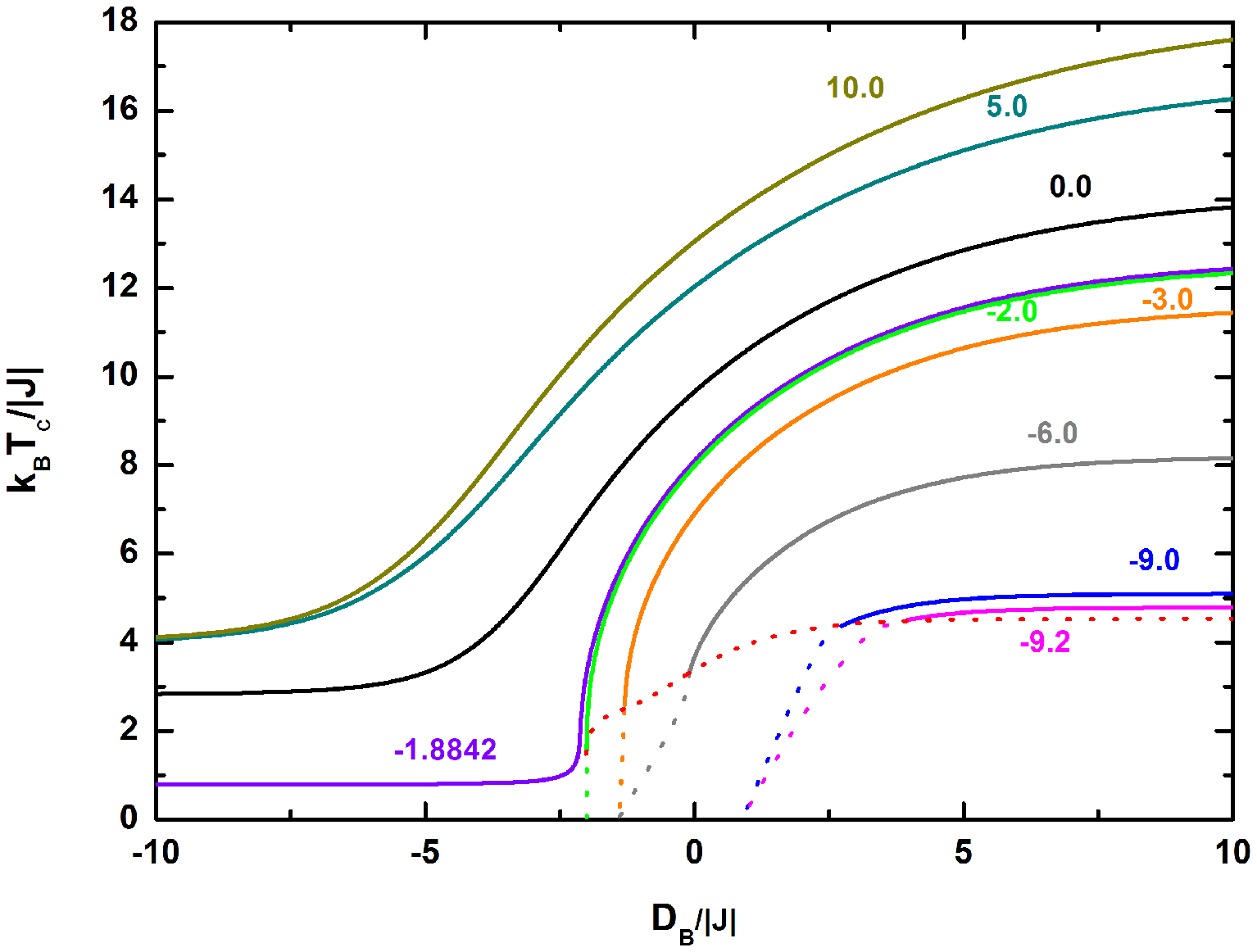} \caption{Phase diagram in the ($D_{B}/\vert J\vert,k_{B}T/\vert J\vert$) plane
for the mixed-spin Ising ferrimagnet with the coordenation number
$z=4$, for several values of $D_{A}/\vert J\vert$. The solid and
light dashed lines, respectively, indicate second- and first-order
phase transitions, while the heavy dashed line represents the positions
of tricritical points. }
\end{figure}

\subsection{Compensation temperature}

The present mixed-spin system can exhibit compensation points, and
to show this fact, we will consider $J<0$. The signs of the sublattice
magnetizations are different since we are taking into account that
in the ferrimagnetic case the system consists of two interpenetrating
square sublattices (A and B) with spin-2 and spin-5/2. Thus, it is
possible that there are a compensation temperature $T_{comp}$ with
$T_{comp}<T_{c}$. So, the total magnetization per site $M=(m_{A}+m_{B})/2$
is equal to zero, same that $m_{A}\not=0$ and $m_{B}\not=0$. 

\begin{figure}[h]
\centering{} \includegraphics[clip,scale=0.44]{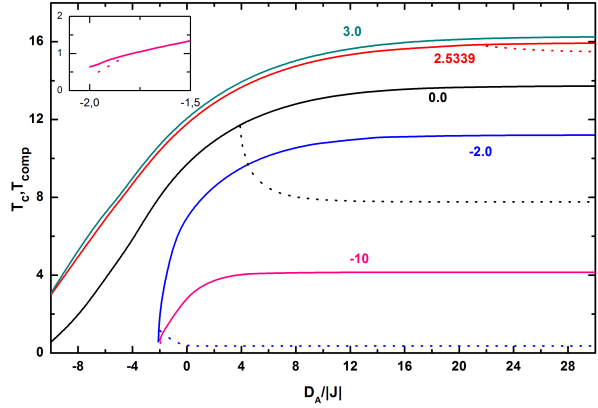} \caption{The critical $T_{c}$ and compensation $T_{comp}$ temperatures as
a function of the single-ion anisotropy $D_{A}/\vert J\vert$, and
for several values of $D_{B}/\vert J\vert$. The solid and dotted
curves represent the critical temperature and compensation temperature,
respectively. The inset shows a magnification of the region closed
where the compensation and critical temperatures are together for
$D_{B}/\left|J\right|=-10.0$. Temperatures are measured in units
of $\left|J\right|/k_{B}$.}
\end{figure}

In Fig. 4, we present the diagram $T_{c}$ and $T_{comp}$ versus
$D_{A}/\vert J\vert$, and for some selected values of $D_{B}/\vert J\vert$.
The diagram shows that there are compensation points in the range
$2.5339<D_{B}/\vert J\vert<-10.0$ and for $D_{A}/\left|J\right|>-2.50$,
which are indicated by dotted lines, while solid lines indicate critical
temperature. The inset in Fig. 4 exhibits some compensation points
in case of $D_{B}/\left|J\right|=-10.0$. The Fig. 5 exhibits the
diagram $T_{c}$ and $T_{comp}$ versus $D_{B}/\vert J\vert$, for
the some selected values of $D_{A}/\vert J\vert$. This figure only
confirms the information shown in Fig. 4, ie, there is compensation
temperature in the range $2.5339<D_{B}/\vert J\vert<-10.0$ and $D_{A}/\left|J\right|>-2.5$. 

\begin{figure}[h]
\centering{} \includegraphics[clip,scale=0.5]{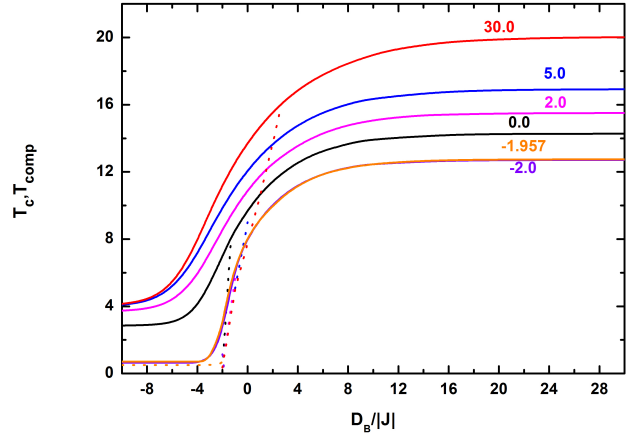} \caption{The critical $T_{c}$ and compensation $T_{comp}$ temperatures as
a function of the single-ion anisotropy $D_{B}/\vert J\vert$, and
for several values of $D_{A}/\vert J\vert$. The solid and dotted
curves represent the critical temperature and compensation temperature,
respectively. Temperatures are measured in units of $\left|J\right|/k_{B}$.}
\end{figure}

\section{Conclusions}

In this paper, we have studied the effects of two different anisotropies
in the phase diagram and in the compensation temperature of the mixed
spin-2 and spin-5/2 ferrimagnetic Ising system by using the mean field
theory based on the Bogoliubov inequality. The phase diagrams are
shown in the critical temperature versus single ions anisotropies
plane. The system presents tricritical behavior, ie, the second-order
phase transition line is separated of the first-order transition line
by a tricritical point. Additionally, we also observed compensation
temperatures. So, in conclusion, we can say that such a system may
exhibit tricritical behavior and compensation temperatures due to
the two different anisotropies.

\section{ACKNOWLEDGMENTS}

This work was supported by the Brazilian Agencies FAPEMAT, CAPES and
CNPq.

\end{document}